# Crystal growth, structural studies and superconducting properties of β-pyrochlore $KOs_2O_6$

G. Schuck[1,2], S. M. Kazakov[1,3], K. Rogacki[1,4], N. D. Zhigadlo[1] and J. Karpinski[1,*]

[1]*Laboratory for Solid State Physics, ETH Zürich, 8093 Zürich, Switzerland*

[2]*present address: Laboratory for Neutron Scattering, PSI, 5232 Villigen, Switzerland*

[3]*present address: Department of Chemistry, Moscow State University, 119899 Moscow, Russia*

[4] *Institute of Low Temperature and Structure Research, PAS, 50-950 Wroclaw, Poland*

**Abstract**

Single crystals of $KOs_2O_6$ have been grown in a sealed quartz ampoule. Detailed single crystal X-ray diffraction studies at room temperature show Bragg peaks that violate $Fd\bar{3}m$ symmetry. With a comparative structure refinement the structure is identified as non-centrosymmetric ($F\bar{4}3m$). Compared to the ideal β-pyrochlore lattice ($Fd\bar{3}m$), both Os tetrahedral and O octahedral network exhibit breathing mode like volume changes accompanied by strong anisotropic character of the K channels. The crystals show metallic conductivity and a sharp transition to the superconducting state at $T_c$ = 9.65 K. Superconducting properties have been investigated by magnetization measurements performed in a temperature range from 2 to 12 K and in magnetic fields from 0 to 60 kOe. The temperature dependence of the upper critical field $H_{c2}(T)$ has been determined and the initial slope $(dH_{c2}/dT)_{Tc}$ = -33.3 kOe/K has been obtained near $T_c$. The upper critical field at zero temperature was estimated to be $H_{c2}(0) \cong 230$ kOe, which is a value close to the Pauli paramagnetic limiting field $H_p(0) \cong 250$ kOe. Then, the Ginzburg-Landau (GL) coherence length $\xi_{GL}(0) \approx 3.8$ nm was calculated, and the Maki parameter $\alpha \approx \sqrt{2}$ was obtained, suggesting the possibility that $KOs_2O_6$ might behave unconventionally at low temperatures and high magnetic fields.

*Corresponding author:

J. Karpinski, karpinski@phys.ethz.ch

## I. Introduction

The discovery of superconductivity in β-pyrochlore oxides $AOs_2O_6$ (A = Cs [1], Rb [2,3] and K [4]) has raised considerable interest. However, until now there is no agreement about the nature of the superconducting pairing mechanism in these compounds. Based on the results of the specific heat measurements,[5] pressure influence on the magnetic field penetration depth,[6] and nuclear magnetic resonance,[7] $RbOs_2O_6$ is suggested to be a weak coupled isotropic BCS superconductor. On the other hand, upper critical field $H_{c2}$ [8], muon spin rotation,[9] and specific heat experiments in $KOs_2O_6$ [10], as well as specific heat experiments in $RbOs_2O_6$ [2] and $CsOs_2O_6$ [11], suggest an unconventional type of paring in these compounds. Most of experiments concerning $KOs_2O_6$ have been performed on polycrystalline samples. The recent availability of single crystals, reported by Hiroi *et al.*[10] and in this article, may help to resolve several issues including the pairing mechanism.

The crystal structure of the β-pyrochlore $\square_2B_2O_6A'$ (B = Os; A' = K, Rb, Cs) can be derived from the ideal structure of α-pyrochlore $A_2B_2O_6O'$, which owns $Fd\bar{3}m$ symmetry (A: 16d; B: 16c; O: 48f and O': 8b).[12] In the ideal β-pyrochlore structure with $Fd\bar{3}m$ symmetry, A' atoms occupy site 8b and the 16d site is empty (Fig. 1). The first discovered α-pyrochlore superconductor $Cd_2Re_2O_7$ shows a transition temperature $T_c$ = 1.5 K [13-15] and, close to 200 K, exhibits a structural phase transition into a tetragonal phase with the non-centrosymmetric space group $I\bar{4}m2$ observed at 13 K.[16,17] Yamaura and Hiroi reported additional Bragg peaks forbidden for $Fd\bar{3}m$ symmetry and breaking the Friedel's Law below 200 K.[18] Also Castellan *et al.* observed additional Bragg peaks and tetragonal splitting below 195 K using high resolution X-ray diffraction.[19] The observation of low-intensity X-ray reflections forbidden for $Fd\bar{3}m$ are also mentioned in numerous publications for other pyrochlore compounds.[20-25]

In this paper we report on the single crystal growth of the $KOs_2O_6$ compound, with a crystallite size up to 300 μm, and on a detailed crystallographic analysis of its crystal structure at room temperature were violation of the ideal $Fd\bar{3}m$ symmetry has been observed. We compare the crystal structures of the β-pyrochlore $KOs_2O_6$ and α-pyrochlore $Cd_2Re_2O_7$ and discuss a possible reason for their different $T_c$'s, We also present the basic superconducting properties of $KOs_2O_6$ in magnetic fields, which show a rather high upper critical field (close to Pauli limiting) and low irreversibility field at low temperatures.

**II. Experimental**

Since osmium easily adopts higher oxidation states and $OsO_4$ is a volatile and toxic compound, closed crucible should be used for the preparation of compounds in the K-Os-O system. Stoichiometric amounts of Os metal (Alfa Aesar, 99.9%) and $KO_2$ (Alfa Aesar, 95%) were thoroughly mixed in an argon filled dry box and pressed into a pellet with a mass of 0.2-0.25 g. The pellet was put into the quartz ampoule with a length of 120 mm, and with an inner and outer diameter of 12 and 14 mm, respectively. A certain amount of silver oxide $Ag_2O$ (Aldrich, 99%) was added to create an appropriate oxygen partial pressure during the synthesis. The ampoule was evacuated and sealed. Then, the ampoule was placed in the preheated furnace at 600 °C, kept for 1 h, cooled down to 400 °C at a rate of 5 °C/h, and finally cooled down to room temperature at a rate of 150 °C/h. As a result, $KOs_2O_6$ single crystals with a size up to 0.3x0.3x0.3 $mm^3$ have been grown on the wall of the quartz ampoule and on the surface of the precursor pellet. The crystals are black and have shapes of octahedra or cubes (Fig. 2).

Structural properties of the $KOs_2O_6$ single crystals have been investigated at room temperature with an x-ray single crystal diffractometer (Bruker SMART CCD system)[26] using Mo K$\alpha$ radiation. 1818 frames were measured with an exposure time of 30 s per frame and $\Delta\Omega = 0.3°$. The cubic unit-cell parameter $a$ = 10.0968(8) Å was refined from 427 reflections. The tetragonal cell parameters were also refined to carry out the comparative structure refinement ($a$ = 7.1392(7) Å, $c$ = 10.0943 (10) Å, tetragonal splitting: $\Delta a$ = 0.002 Å, $c/a$ = 1.0002). The three-dimensional data were reduced and corrected for Lorentz, polarization and background effects using SAINT and corrected for absorption using SADABS.[26] Due to strong influence of the absorption effect two-third of the measured data shows significant scaling fluctuations after initial absorption correction. Thus, we finally use only that part of the data showing a flat scaling behavior. The structure was solved by direct methods and difference Fourier synthesis using SHELXS-97 and the fullmatrix least-squares refinement on $F^2$ was achieved with SHELXL-97.[27,28] All reciprocal space scans were performed with a Siemens P4-four circle diffractometer with Mo K$\alpha$ radiation. The results of X-ray measurements and parameters of the structure refinement are given in Table I.

Superconducting properties of the $KOs_2O_6$ single crystals have been investigated by magnetization measurements performed as a function of temperature and field on a homemade SQUID magnetometer with a Quantum Design sensor. Individual crystals as well as a collection of 28 crystals with a mass of 887 μg were studied to obtain reliable quantitative results. Magnetization, M, was measured at constant field upon heating from zero-field-cooled state (ZFC mode) and upon cooling (FC mode). The temperature sweep was about 0.2 K/min resulting in accuracy better than $5 \cdot 10^{-7}$ emu for an option where results of 6 scans were averaged to obtain each individual experimental point.

**III. Crystal structure**

The observation of low-intensity X-ray reflections forbidden in the $Fd\bar{3}m$ space group or other symmetry violations have been reported for numerous pyrochlore compounds.[17-25] Not all types of symmetry violations known in literature have been found for $KOs_2O_6$ using our data obtained for single crystals. No clear indications were received that the Friedel`s law is broken. However, five Bragg reflections that are inconsistent with d-glide symmetry (0kl: k+l ≠ 4n: 024, 046, 028, 041 and 0212) and three Bragg reflections that are inconsistent with $4_1$ screw axis (00l: l ≠ 4n: 002, 006 and 0010) of $Fd\bar{3}m$ were measured with the CCD system (I > 2.00 σ(I)). Bragg peaks violating the $Fd\bar{3}m$ symmetry were also observed in reciprocal space scans (Fig. 3), while no Bragg peaks violating the F-centering were found with similar scans. Careful analyses of the reciprocal space scans of the 0 0 12 and 0 0 16 Bragg peaks were performed to see possible tetragonal splitting. No clear evidence for such splitting was observed.

Two non-centrosymmetric space groups $F\bar{4}3m$ and F23 were used successfully instead of $Fd\bar{3}m$ to describe the lowered symmetry in β-pyrochlore $KOs_2O_6$. For the comparative structure refinement, centrosymmetric $Fd\bar{3}m$ and two tetragonal non-centrosymmetric $I\bar{4}m2$ and $I4_122$ space groups (proposed for the low temperature phases of $Cd_2Re_2O_7$)[18] were also taken into consideration. The results of these refinements are summarized in Table I. Direct methods have been used to locate the Os atom position for both cubic space groups: $Fd\bar{3}m$ (16c) and $F\bar{4}3m$ (16e). Refined atomic parameters, anisotropic (ADP) and isotropic (IDP) displacement parameters for the used cubic space groups are listed in Table II. Selected interatomic distances are given in Table III.

Using $Fd\bar{3}m$ and 16c site for Os (fixed IDP) yielded a residual of 0.15. The difference Fourier synthesis revealed two strong peaks for K at 8b and for O at 48f. Introducing ADP's for K and IDP's for Os and O reduced the residual to $R_1 = 0.0309$ (Table I). Initial refinement with $F\bar{4}3m$ and 16e site for Os yields to a residual of 0.12 using fixed IDP's for the Os. The positions for K and O have been derived form the $Fd\bar{3}m$ refinement. The K and O atom positions in $Fd\bar{3}m$ space group split in $F\bar{4}3m$ symmetry each into two sites (K: K1 at 4c site and K2 at 4b site and O: O1 at 24f site and O2 at 24g site). The refinement with ADP's for K1 and K2 atoms and with IDP's for Os, O1 and O2 (using the same IDP's for O1 and O2) yielded a residual of $R_1 = 0.0285$ (Tab. I). For F23 the structural parameters are basically the same like for $F\bar{4}3m$ except for additional degree of freedom for the ADP's of the O1 and the O2 atoms ($U_{22} \neq U_{33}$). The final refinement with F23 includes ADP's for K1 and K2 and IDP's for Os, O1 and O2 (using the same IDP's for O1 and O2) (Table II). Simultaneous refinement of the occupancies and refinement of ADP's for O and Os was not possible for all considered space groups.

The differences in residuals resulted from the comparative structure refinement are small, however the deviation from ideal β-pyrochlore is evident. The tetragonal space group symmetries exhibit higher residuals so that we conclude that β-pyrochlore $KOs_2O_6$ at room temperature is cubic. This is even more evident taking into account the fact that the tetragonal splitting could not be confirmed in the Bragg peak profile analysis. Even though there are minor differences between the both non-centrosymmetric cubic space groups, $F\bar{4}3m$ was chosen over F23 because of lower residuals and smaller residual electron density values.

The crystal structure of β-pyrochlore $KOs_2O_6$, described with $F\bar{4}3m$, can be considered as based on $OsO_6$ octahedra linked trough all vertices. The $OsO_6$ octahedra form channels whereas the K atoms K1 and K2 are located on opposite sites of the channels (Fig. 1). Compared to the ideal β-pyrochlore structure (short O-K bond lengths of 3.099(5) Å), one shortest O-K distance (O1-K2) decreases to 3.059(13) Å and the other shortest O-K distance (O2-K1) increases to 3.141(13) Å. The difference between the both shortest O-K distances is 0.088 Å, what indicates a strong anisotropic character of the K channels where the K atom on the K1 position has more freedom to move (Tab. 3 and Fig. 4). Additionally to the anisotropic character of the K channels, large atomic displacement parameters of the K atoms ($U_{iso}$ = 0.06 Å$^3$) are observed compared to those of the O atoms ($U_{iso}$ = 0.005 Å$^3$) and even more to the Os atoms ($U_{iso}$ = 0.0005 Å$^3$). These observations correlate well with the results of band structure calculations of Kuneš *et al.* where they found that the K$^+$ ion optic mode is unstable giving rise to a 'rattling' behaviour of the K atoms.[29]

The symmetry lowering leads to even more anisotropic features in the structure of β-pyrochlore KOs$_2$O$_6$. The strongest effect is observed on O-O distances in the octahedral network of KOs$_2$O$_6$. Instead of the identical O-O distance in the empty O$_6$ octahedra in the ideal $Fd\bar{3}m$ structure there are two different O-O distances observed with $F\bar{4}3m$ symmetry, with a difference of 0.124 Å (~ 4%) between them (Fig. 5). Compared to the ideal structure (O-O = 2.757 Å), one O-O distance increases to 2.819 Å (O1-O1) and the other O-O distances decreases to 2.695 Å (O2-O2). This gives rise to a breathing-mode like behaviour of the O$_6$ octahedra with two different octahedra volumes (Fig. 5). A similar effect with two different octahedra volumes but with four different O-O distances was reported by Weller *et al.* for the low temperature phase of the α-pyrochlore Cd$_2$Re$_2$O$_7$, with a maximal difference of the O-O distances of about 0.31 Å (~11%).[17] The O-O distances of the OsO$_6$ octahedra (2.653 Å) in the β-pyrochlore KOs$_2$O$_6$ do not change compared to the ideal symmetry (Table 3). The Os tetrahedral network of β-pyrochlore KOs$_2$O$_6$ also shows a breathing mode like behaviour similar to the low temperature phase of α-pyrochlore Cd$_2$Re$_2$O$_7$ (Fig. 6).

**IV. Superconducting properties**

In Fig. 7 we show the magnetization curves of the KOs$_2$O$_6$ single crystals measured with increasing (ZFC mode) and decreasing (FC mode) temperature. A sharp transition to the superconducting state is observed at $T_c$ = 9.65 K in 10 Oe magnetic field (see inset). This transition broadens significantly with increasing H, indicative weak pinning properties and, consequently, a low irreversibility field $H_{irr}$. For example, at a temperature of 9.2 K the upper critical field $H_{c2}$ = 15 kOe, however $H_{irr}$ is equal only to about 0.5 kOe. At lower temperatures $H_{irr}$ does not exceed a rather low value of 10 kOe, for the critical current criterion $j_c \approx 6$ A/cm$^2$ corresponding to the magnetization resolution $\Delta M \approx 5 \cdot 10^{-7}$ emu. Such low $H_{irr}$ reveals marginal importance of KOs$_2$O$_6$ as a material for high current applications, even if we assume that the crystals are extremely clean and thus the pinning properties of this compound can be improved significantly in polycrystalline materials. On the other hand, the upper critical field is expected to be extremely high at low temperatures ($H_{c2}(0) \cong 230$ kOe), pointing to unusual properties of $KOs_2O_6$. We discuss this in the paragraphs devoted to the upper critical field.

Magnetization curves M(T) similar to those presented in Fig. 7 have been measured at fields up to 60 kOe to obtain the temperature dependence of the upper critical field $H_{c2}(T)$. In Fig. 8 we show $H_{c2}(T)$ near $T_c$ revealing a linear dependence. This linear dependence is obtained if a "bulk" criterion, i.e. a bend in the M(T) curve (see Fig. 7) is used to determine $T_c$ at a given field. However, at low fields a weak diamagnetic signal appears about 0.05-0.1 K above a bulk transition at $T_c$. This weak diamagnetism disappears at higher fields, thus seems to be a surface effect, and it may be responsible for a slightly upward curvature of $H_{c2}(T)$ observed from resistivity measurements for single-crystalline $KOs_2O_6$ blocks.[10] In our work, special attention has been paid to determine $H_{c2}(T)$ in the vicinity of $T_c$. The initial slope $(-dH_{c2}/dT)_{Tc} = 33.3$ kOe/K has been obtained for the $KOs_2O_6$ single crystals. This value is similar to 34 kOe/K reported for both single-crystalline and polycrystalline samples.[10,30] The similar initial slope of $H_{c2}(T)$, observed for the single crystals and polycrystalline materials, suggests that the scattering of electrons by structural imperfections or impurities remains essentially unaffected.

Magnetization of the $KOs_2O_6$ crystals was measured as a function of field to show basic pinning properties and to estimate the lower critical field, $H_{c1}$, at low temperatures. As an example, in Fig. 8 (inset) we show a virgin (ZFC) magnetization half-loop obtained at 5 K for H swept from zero to 5 kOe and then back to zero. At T = 5 K the irreversibility field $H_{irr} \approx$ 10 kOe is unexpectedly low and the M(H) half-loop is surprisingly narrow if consider that $H_{c2}(0) \approx 200$ kOe is expected for these crystals, as we show in next paragraphs. For fields higher than 3 kOe, the half-loop is nearly reversible and this feature, together with low $H_{irr}$, confirms a weak pinning force revealed also in magnetization measurements performed as a function of temperature at various fields (see Fig. 7). In Fig. 8 (inset) the linear part of the M(H) half-loop (Meissner state) is used to evaluate the demagnetizing factor N = 0.52 by assuming perfect diamagnetism. This value seems to be reasonable for randomly oriented octahedral-shaped crystals, since N = 0.33 is the demagnetizing factor for a sphere. Then, the field where a deviation from a linear M(H) dependence begins to appear is taken as the first penetration field, $H_{fp}$, which is an upper limit for the thermodynamic lower critical field $H_{c1}$. After correction for demagnetizing effects (N = 0.52), $H_{fp}$(5 K) ≅ 95 Oe is obtained and $H_{c1}$(5 K) ≈ 90 Oe is estimated, considering that the crystals are characterized by a rather low surface barrier, as usually observed for a case different from a plane surface oriented parallel to the external field.

The orbital upper critical field at zero temperature, $H_{c2}^*(0)$, can be calculated in the frame of the Ginzburg-Landau-Abrikosov-Gor'kov (GLAG) theory from a simple formula: $H_{c2}^*(0) = -k(dH_{c2}/dT)_{Tc} \cdot T_c$, where k is a constant equal to 0.69 and 0.73, for the dirty and clean limits, respectively. Thus, $H_{c2}^*(0)$ = 222 kOe (dirty) and 235 kOe (clean) has been obtained for the $KOs_2O_6$ crystals. These values have been derived with an assumption that the spin-orbit scattering parameter $\lambda_{so} = \infty$, thus the superconducting pairs are broken due to the orbital effects only. Strong spin-orbit scattering seems to be required for an interpretation of the high superconducting transition temperature of the $KOs_2O_6$ compound, as it results from the electronic structure analyses.[29] However, the Pauli paramagnetic limiting field in the BCS Clogston approach is $H_p^{BCS}(0) = 18.4 \cdot T_c$ = 178 kOe and, consequently, the Maki parameter $\alpha^{BCS} = \sqrt{2} \cdot H_{c2}^*(0)/H_p^{BCS}(0) \cong 1.9$ (clean limit). This suggests that the spin contribution to the pair-breaking effect is significant.[31-33] Thus, $H_{c2}$ in $KOs_2O_6$ seems to be influenced by the Pauli paramagnetic effect and the first order transition from the superconducting to the normal state is possible at low temperatures. This suggests that the quantum critical state and the Fulde-Ferrell-Larkin-Ovchinnikov (FFLO) phase with the spatially modulated superconducting order parameter may appear at low temperatures and high magnetic fields.[34-37] We note, however, that an enhancement of the BCS Pauli limiting field is expected when the electron-phonon interaction is strong.[38] This seems to be the case for $KOs_2O_6$ where an electron-phonon coupling parameter $\lambda_{eph} \approx 1$ has been derived from specific heat measurements.[39] Thus, the enhanced Pauli limiting field is $H_p(0) = H_p^{BCS}(0)\sqrt{(1 + \lambda_{eph})} \approx 250$ kOe and, consequently, the Maki parameter $\alpha$ = 1.33 ($\approx \sqrt{2}$). This result shows that even if the strong electron-phonon coupling is considered, $H_p(0)$ is comparable to $H_{c2}^*(0)$ and unusual critical state phenomena may occur for $KOs_2O_6$ in high magnetic fields. Moreover, in this pure compound where $H_{c2}^*(0) \sim H_p(0)$, the hypothetical type-IV superconductivity with the Cooper pairs with broken inversion and time-reversal symmetries may also exist.[40]

Encouraged by the Maki parameter $\alpha \sim \sqrt{2}$ we estimate a reasonable value of the upper critical field at low temperatures $H_{c2}(0) \cong 230$ kOe, as a value of the orbital critical field $H_{c2}^*(0)$ not influenced by paramagnetic effects ($H_p \cong 250$ kOe). This $H_{c2}(0)$ can be used to evaluate the Ginzburg-Landau (GL) coherence length at zero temperature $\xi_{GL}(0) = (\Phi_0/2\pi H_{c2})^{1/2} \cong 3.8$ nm, where $\Phi_o$ is the magnetic flux quantum. Then, the GL parameter can be estimated $\kappa_{GL} = \lambda_{GL}(0)/\xi_{GL}(0) \cong 70$, with the GL penetration depth $\lambda_{GL}(0) \cong 270$ nm obtained in the recent muon-spin-rotation experiments.[8] The large $\kappa_{GL}$ value classifies $KOs_2O_6$ as a strong type-II superconductor and makes it possible to evaluate the lower critical field by using a formula $H_{c1}(0) = \Phi_0 \ln\kappa/4\pi\lambda^2 \cong 96$ Oe. This field is only slightly higher than $H_{c1}(5\,K) \approx 90$ Oe derived from the M(H) measurements and thus seems to confirm reliability of our analysis performed in the frame of the GLAG theory.

The initial slope of the upper critical field $(-dH_{c2}/dT)_{Tc} = 33.3$ kOe/K obtained for $KOs_2O_6$ is much larger than those reported for other β-pyrochlore members, 17 and 12 kOe/K for $RbOs_2O_6$ and $CsOs_2O_6$, respectively.[5,9] This points to $KOs_2O_6$ as to a superconductor with thermodynamic properties much different than those observed for Rb and Cs substituted counterparts. Therefore, more unconventional mechanism than the phonon-mediated s-wave pairing may be considered as a possible solution for the $KOs_2O_6$ compound. The large increase of $(-dH_{c2}/dT)$ with $T_c$ increased across the Cs-Rb-K series requires a considerable enhancement of the scattering rate of charge carriers and, in consequence, a reduction of the GL coherence length. Such an increase of the scattering rate of carriers seems to be reflected in peculiar downward curvature of the temperature dependence of the resistivity observed for $KOs_2O_6$.[9] A substantial reduction of $\xi_{GL}(0)$ from 7.4 nm reported for $RbOs_2O_6$ [Ref. 5] to 3.8 nm obtained for $KOs_2O_6$ is also observed. This supports the scenario where an enhancement of the scattering of carriers due to the unusual low-energy dynamics and increased anharmonicity of the $K^+$ ions is considered as an effect responsible for the high superconducting transition temperature of the $KOs_2O_6$ compound.[29] The unusual low-energy dynamics and associated electron-phonon coupling can be also considered as the cause of quantitative differences observed in physical properties of $KOs_2O_6$ compared to other β-pyrochlores. In order to clarify this question, the K-ion dynamics should be studied in detail, e.g., by NMR experiments,[41] and the reliable electron-phonon coupling strength might be extracted from the high-resolution angle-resolved photoemission spectroscopy.[42]

## V. Conclusions

High quality single crystals of the β-pyrochlore $KOs_2O_6$ superconductor have been grown and the crystal structure and basic superconducting properties have been studied in detail. With a comparative structure refinement and careful Bragg peak analysis we have received clear evidence that the previously proposed ideal centrosymmetric $Fd\bar{3}m$ space group symmetry is broken at room temperature. The best description of the crystal structure has been achieved with the non-centrosymmetric $F\bar{4}3m$ space group. The most striking feature of this symmetry lowering is the breathing mode like behaviour of the octahedral and tetrahedral networks accompanied by drastic bond length changes compared to the ideal structure. The cubic crystal structure of β-pyrochlore $KOs_2O_6$, refined at room temperature, exhibits several similarities to the tetragonal low temperature phase of α-pyrochlore $Cd_2Re_2O_7$.[17] Thus, higher $T_c$ of $KOs_2O_6$ may be related to its lower symmetry. Besides the similarities between $KOs_2O_6$ and $Cd_2Re_2O_7$, the anisotropic character of the K channels is a notable difference. Moreover, band structure calculations show significant instability and anharmonicity of the K ions ("rattling"),[29] which seems to be consistent with a rather high anisotropic displacement parameter derived for the K ions from our crystal structure studies.

Basic superconducting properties of $KOs_2O_6$ single crystals have been determined by magnetization measurements. These properties show that $KOs_2O_6$ is a type-II superconductor with a rather large Ginzburg-Landau parameter $\kappa_{GL} \cong 70$. The crystals reveal very weak pinning properties leading to a low irreversibility field and making this material less important for any high-current high-field application. The upper critical field at low temperatures is surprisingly high, $H_{c2}(0) \cong 230$ kOe, and comparable to the paramagnetic limiting field $H_p \cong 250$ kOe. Thus, $KOs_2O_6$ seems to be a rather unconventional superconductor, where several exotic superconducting states may occur at high magnetic fields. One is the Fulde-Ferrell-Larkin-Ovchinnikov phase with the spatially modulated superconducting order parameter. The other state is the superconducting phase where a mixture of the Cooper pairs with the singlet and triplet symmetries has been predicted to exist.[40] Advanced spectroscopic studies in fields of about 250 kOe are required to verify these intriguing hypotheses.

**Acknowledgements**

This work has been supported by NCCR MaNEP and National Science Foundation. The SMART CCD measurements were performed at the Laboratory of Inorganic Chemistry, ETH Zürich. We are grateful to M. Wörle (Laboratory of Inorganic Chemistry, ETH Zürich) for fruitful discussion. We also acknowledge experimental help of R. Puzniak.

**Note from authors**

During resubmitting our paper we noticed, that J. Yamaura et al. reported the crystal structure of $KOs_2O_6$ based on a space group $Fd\bar{3}m$ [43].

**References**

1. S. Yonezawa, Y. Muraoka and Z. Hiroi: J. Phys. Soc. Jpn. **73** (2004) 1655.
2. S. Yonezawa, Y. Muraoka, Y. Matsushita and Z. Hiroi: J. Phys. Soc. Jpn. **73** (2004) 819. Results obtained in magnetic fields should be corrected according to the erratum published in; Z. Hiroi, S. Yonezawa, J. Yamaura, T. Muramatsu, Y. Matsushita, and Y. Muraoka: J. Phys. Soc. Jpn. **74** (2005) 3400.
3. S. M. Kazakov, N. D. Zhigadlo, M. Brühwiler, B. Batlogg and J. Karpinski: Supercond. Sci. Technol. **17** (2004) 1169.
4. S. Yonezawa, Y. Muraoka, Y. Matsushita and Z. Hiroi: J. Phys.: Condens. Matter **16** (2004) L9.
5. M. Brühwiler, S. M. Kazakov, N. D. Zhigadlo, J. Karpinski and B. Batlogg: Phys. Rev. B **70** (2004) 020503R.
6. R. Khasanov, D. G. Eshchenko, J. Karpinski, S. M. Kazakov, N. D. Zhigadlo, R. Brütsch, D. Gavillet, D. Di Castro, A. Shengelaya, F. La Mattina, A. Maisuradze, C. Baines, and H. Keller, Phys. Rev. Lett. **93**, 157004 (2004).
7. K. Magishi, J. L. Gavilano, B. Pedrini, J. Hinderer, M. Weller, H. R. Ott, S. M. Kazakov, and J. Karpinski, Phys. Rev. B **71**, 024524 (2005).
8. Z. Hiroi, S. Yonezawa, Y. Muraoka, J. Phys. Soc. Jpn. **73** (2004) 1651.
9. A. Koda, W. Higemoto, K. Ohishi, S. R. Saha, R. Kadono, S. Yonezawa, Y. Muraoka, and Z. Hiroi, J. Phys. Soc. Jpn. **74** (2005) 1678.
10. Z. Hiroi, S. Yonezawa, J. Yamaura, T. Muramatsu, and Y. Muraoka, J. Phys. Soc. Jpn. **74** (2005) 1682. Results obtained in magnetic fields should be corrected according to the erratum published in; Z. Hiroi, S. Yonezawa, J. Yamaura, T. Muramatsu, Y. Matsushita, and Y. Muraoka: J. Phys. Soc. Jpn. **74** (2005) 3400 .
11. Z. Hiroi, S. Yonezawa, T. Muramatsu, J. Yamaura, and Y. Muraoka, J. Phys. Soc. Jpn. **74** (2005) 1255. Results obtained in magnetic fields should be corrected according to the erratum published in; Z. Hiroi, S. Yonezawa, J. Yamaura, T. Muramatsu, Y. Matsushita, and Y. Muraoka: J. Phys. Soc. Jpn. **74** (2005) 3400.
12. M.A. Subramanian, G. Aravamudan, and G.V.S. Rao, Prog. Solid Sate Chem. **15**, 55 (1983).

13. M. Hanawa, Y. Muraoka, T. Tayama, T. Sakakibara, J. Yamaura and Z. Hiroi, Phys. Rev Lett. **87**, 187001 (2001).
14. H. Sakai, K. Yoshimura, H. Ohno, H. Kato, S. Kambe, R. E. Walstedt, T. D. Matsuda, Y. Haga and Y. Onuki, J. Phys.: Condens. Matter, **13**, L785 (2001).
15. R. Jin, J. He, S. McCall, C. S. Alexander, F. Drymiotis and D. Mandrus, Phys. Rev. B, **64**, 180503 (2001).
16. Z. Hiroi, Y. Yamaura, Y. Muraoka and M. Hanawa, , J. Phys. Soc. Jpn., **71**, 1634 (2002).
17. M. T. Weller, R. H. Hughes, J. Rooke, C. S. Knee and J. Reading, Dalton. Trans., 3032 (2004).
18. J. Yamaura and Z. Hiroi, J. Phys. Soc. Jpn., **71**, 2598 (2002).
19. J. P. Castellan, B. D. Gaulin, J. van Duijn, M. J. Lewis, M.D Lumsden, R. Jin, J. He, S. E. Nagler and D. Mandrus, Phys. Rev. B **66**, 134528 (2002).
20. R. C. Rouse, P. J. Dunn, D. R. Peacor and L. Wang, J. Solid State Chem. **141**, 562 (1998).
21. B. J. Kennedy, J. Solid State Chem. **123**, 14 (1996).
22. A. W. Sleight, F. C. Zumsteg, J. R. Barkley, and J. E. Gulley, Mat. Res. Bull. **13**, 1247 (1978).
23. H. Kobayashi, R. Kanno, Y. Kawamoto, T. Kamiyama, F. Izumi, and A. W. Sleight, J. Solid State Chem. **114,** 15 (1995).
24. C. Michel, D. Groult, and B. Raveau, J. Inorg. Nucl. Chem. **37**, 247 (1975).
25. R. A. Beyerlein, H. S. Horowitz, J. M. Longo and M. E. Leonowicz, J. Solid State Chem. **51**, 253 (1984).
26. Bruker (2002). *SMART* (Version 5.62), *SAINT* (Version 6.02) and *SADABS* (Version 2.03) Bruker AXS Inc., Madison, Wisconsin, USA.
27. SHELX97 [Includes SHELXS97, SHELXL97] - Programs for Crystal Structure Analysis (Release 97-2). G. M. Sheldrick, Institüt für Anorganische Chemie der Universität, Tammanstrasse 4, D-3400 Göttingen, Germany, 1998.
28. L. J. Farrugia, J. Appl. Cryst. **32**, 837 (1999).
29. J. Kuneš, T. Jeong and W.E. Pickett, Phys. Rev. B **70**, 174510 (2004).

30. Z. Hiroi, S. Yonezawa, and Y. Muraoka, J. Phys. Soc. Jpn. **73**, 1651 (2004). Results obtained in magnetic fields should be corrected according to the erratum published in; Z. Hiroi, S. Yonezawa, and Y. Muraoka, J. Phys. Soc. Jpn. **74** (2005) 3399.
31. T. P. Orlando, E. J. McNiff, S. Foner, and M. R. Beasley, Phys. Rev. B **19**, 4545 (1979).
32. K. Maki, Physica **1**, 127 (1964).
33. N. R. Werthamer, E. Helfand, and P. C. Hohenberg, Phys. Rev. **147**, 295 (1966).
34. P. Fulde and R. A. Ferrell, Phys. Rev. **135**, A550 (1964).
35. A. I. Larkin and Y. N. Ovchinnikov, ZhETF **47**, 1136 (1964) [Sov. Phys. JETP **20**, 762 (1965)].
36. K. Maki, Phys. Rev. **148**, 362 (1966).
37. L. Gruenberg and L. Gunther, Phys. Rev. Lett. **16**, 996 (1966).
38. T. P. Orlando and M. R. Beasley, Phys. Rev. Lett. **46**, 1598 (1981).
39. M. Brühwiler, S. M. Kazakov, J. Karpinski, and B. Batlogg, Phys. Rev. B, submitted.
40. A. G. Lebed, ZhETF **82**, 223 (2005) [Sov. Phys. JETP **82**, 204 (2005)], and references cited therein.
41. K. Arai, J. Kikuchi, K. Kodama, M. Takigawa, S. Yonezawa, Y. Muraoka, and Z. Hiroi, cond/mat 0411460.
42. S.-J. Tang, Junren Shi, Biao Wu, P.T. Sprunger, W.L. Yang, V. Brouet, X.J. Zhou, Z. Hussain, Z.-X. Shen, Zhenyu Zhang, and E.W. Plummer, Phys. Stat. Sol. (b) **241**, 2345 (2004)
43. J. Yamaura, S. Yonezawa, Y. Muraoka, Z. Hiroi, Journal of Solid State Chemistry **179,** 334 (2005).
44. H. D. Flack, Acta Cryst. **A39**, 876 (1983).

TABLE I. Crystallographic and structure refinement parameters of β-pyrochlore $KOs_2O_6$ at room temperature. The refinements were performed with CCD data (Mo K$\alpha$ / $\lambda$ = 0.71073 Å). Calculated $\rho$ and $\mu$: $\rho$ = 6.653 g/cm$^3$ and $\mu$ = 50.087 mm$^{-1}$. The crystal size was ~0.1 mm$^3$.

| Crystal system | cubic | | | tetragonal | |
|---|---|---|---|---|---|
| Unit cell *a* (Å) | 10.0968 (9) | | | 7.1392(7) | |
| Unit cell *c* (Å) | 10.0968 (9) | | | 10.0943(10) | |
| Volume (Å$^3$) | 1029.32 (16) | | | 514.49(9) | |
| F$_{000}$ | 1752 | | | 876 | |
| Z | 8 | | | 4 | |
| θ range (deg) | 3.49 - 30.18 | | | 3.50 - 33.67 | |
| h$_{min}$; k$_{min}$; l$_{min}$ | -13; -14; -6 | | | -10; -6; -15 | |
| h$_{max}$; k$_{max}$; l$_{max}$ | 7; 14; 14 | | | 2; 5; 14 | |
| **Space group** | **Fd$\bar{3}$m** | **F$\bar{4}$3m** | **F23** | **I$\bar{4}$m2** | **I4$_1$22** |
| Inversion symmetry | yes | no | no | no | no |
| Extinction | 0.00282(4) | 0.0029(1) | 0.0030(1) | 0.0125(2) | 0.0126(2) |
| Flack parameter [44] | / | 0.4(2) | 0.47(19) | 0.49(13) | 0.56(14) |
| Total Ref. | 1080 | 1170 | 1170 | 1331 | 1326 |
| Observations | 98 | 194 | 274 | 531 | 473 |
| Variables | 6 | 10 | 10 | 15 | 12 |
| R$_{int}$ | 0.0285 | 0.0270 | 0.0267 | 0.0301 | 0.0308 |
| R$_{sig}$ | 0.0152 | 0.0204 | 0.0216 | 0.0354 | 0.0325 |
| R$_1$ (observed data) | 0.0309 | 0.0285 | 0.0299 | 0.0330 | 0.0347 |
| wR$_2$ (observed data) | 0.0668 | 0.0661 | 0.0677 | 0.0744 | 0.0747 |
| S (observed data) | 1.379 | 1.325 | 1.331 | 1.210 | 1.284 |
| Largest Δρ (x,y,z), max; min (e/Å$^3$) | 1.583; -1.911 | 1.742; -2.413 | 2.032; -2.536 | 2.753; -2.882 | 2.479; -2.891 |

TABLE II Atomic coordinates, anisotropic ($U_{ij}$) and isotropic displacement parameters ($U_{iso}$) in Å$^3$ of β-pyrochlore KOs$_2$O$_6$ at room temperature. The results were obtained from the comparative structure refinement by using cubic and tetragonal space groups.

| Site | | x | y | z | $U_{11}$ | $U_{22}$ | $U_{33}$ | $U_{12}$ | $U_{13}$ | $U_{23}$ | $U_{iso}$ |
|---|---|---|---|---|---|---|---|---|---|---|---|
| **Space group: Fd$\bar{3}$m (No. 227; setting 2)** | | | | | | | | | | | |
| Os | 16c | 0 | 0 | 0 | - | - | - | - | - | - | 0.0005(1) |
| K | 8b | 3/8 | 3/8 | 3/8 | 0.064(3) | $U_{11}$ | $U_{11}$ | 0 | 0 | 0 | 0.064(3) |
| O | 48f | 0.3181(5) | 1/8 | 1/8 | - | - | - | - | - | - | 0.0041(8) |
| **Space group: F$\bar{4}$3m (No. 216)** | | | | | | | | | | | |
| Os | 16e | 0.8758(1) | x | x | - | - | - | - | - | - | 0.0005(1) |
| K1 | 4c | 1/4 | 1/4 | 1/4 | 0.063(5) | $U_{11}$ | $U_{11}$ | 0 | 0 | 0 | 0.063(5) |
| K2 | 4b | 1/2 | 1/2 | 1/2 | 0.062(4) | $U_{11}$ | $U_{11}$ | 0 | 0 | 0 | 0.062(4) |
| O1 | 24f | 0.197(1) | 0 | 0 | - | - | - | - | - | - | 0.005(1) |
| O2 | 24g | 0.561(1) | 1/4 | 1/4 | - | - | - | - | - | - | 0.005(1) |
| **Space group: F23 (No. 196)** | | | | | | | | | | | |
| Os | 16e | 0.8757(1) | x | x | - | - | - | - | - | - | 0.0006(1) |
| K1 | 4c | 1/4 | 1/4 | 1/4 | 0.089(8) | $U_{11}$ | $U_{11}$ | 0 | 0 | 0 | 0.089(8) |
| K2 | 4b | 1/2 | 1/2 | 1/2 | 0.044(3) | $U_{11}$ | $U_{11}$ | 0 | 0 | 0 | 0.044(3) |
| O1 | 24f | 0.197(1) | 0 | 0 | - | - | - | - | - | - | 0.005(1) |
| O2 | 24g | 0.561(1) | 1/4 | 1/4 | - | - | - | - | - | - | 0.005(1) |
| **Space group: I$\bar{4}$m2 (No. 119)** | | | | | | | | | | | |
| Os | | 0.2485(1) | 0 | 0.8743(1) | - | - | - | - | - | - | 0.0009(1) |
| K1 | | 0 | 0 | 1/2 | 0.082(8) | $U_{11}$ | 0.08(1) | 0 | 0 | 0 | 0.082(6) |
| K2 | | 0 | 1/2 | 3/4 | 0.048(4) | $U_{11}$ | 0.06(1) | 0 | 0 | 0 | 0.051(3) |
| O1 | | 0.1843(9) | x | 0 | - | - | - | - | - | - | 0.0045(7) |
| O2 | | 0.199(1) | 0.699(1) | 1/4 | - | - | - | - | - | - | 0.0045(7) |
| O3 | | 0 | 0 | 0.200(1) | - | - | - | - | - | - | 0.0045(7) |
| O4 | | 0 | 1/2 | 0.432(1) | - | - | - | - | - | - | 0.0045(7) |
| **Space group: I4$_1$22 (No. 98)** | | | | | | | | | | | |
| Os | | 1/4 | 0.0013(1) | 7/8 | - | - | - | - | - | - | 0.0009(1) |
| K1 | | 0 | 0 | 1/2 | 0.065(3) | $U_{11}$ | 0.06(1) | 0 | 0 | 0.03(1) | 0.066(2) |
| O1 | | 0.1808(9) | x | 0 | - | - | - | - | - | - | 0.0035(7) |
| O2 | | 0.7975(9) | -x | 0 | - | - | - | - | - | - | 0.0035(7) |
| O3 | | 0 | 0 | 0.1913(7) | - | - | - | - | - | - | 0.0035(7) |

TABLE III Selected interatomic distances (Å) and bond angles (°) of β-pyrochlore $KOs_2O_6$.

| Space group | | $Fd\bar{3}m$ | $F\bar{4}3m$ | $F23$ |
|---|---|---|---|---|
| **osmium environment** | | | | |
| Os – O | [6]*[1] | 1.913(2) | - | - |
| Os – O1 | [3] | - | 1.920(5) | 1.921(4) |
| Os – O2 | [3] | - | 1.906(4) | 1.905(4) |
| O – Os – O | | 87.8(2) | - | - |
| O – Os – O | | 92.2(2) | - | - |
| O – Os – O | | 180.0(3) | - | - |
| Os – O – Os | | 138.0(3) | - | - |
| O1 – Os – O1 | | - | 94.2(5) | 94.3(4) |
| O1 – Os – O2 | | - | 87.8(3) | 87.8(3) |
| O2 – Os – O2 | | - | 90.1(5) | 90.1(4) |
| O1 – Os – O2 | | - | 177.0(6) | 177.0(5) |
| Os – O1 – Os | | - | 137.7(3) | 138.0(3) |
| Os – O2 – Os | | - | 138.9(6) | 138.9(4) |
| **tetrahedral network** | | | | |
| Os – Os | [8] | 3.570(4) | - | - |
| Os – Os | [4] | - | 3.548(4) | 3.549(4) |
| Os – Os | [4] | - | 3.592(5) | 3.591(5) |
| **potassium environment** | | | | |
| K – O | [8] | 3.099(5) | - | - |
| K1 – O2 | [8] | - | 3.141(13) | 3.142(11) |
| K2 – O1 | [8] | - | 3.059(13) | 3.057(11) |
| K – K | [4] | 4.3720(4) | - | - |
| K1 – K2 | [4] | - | 4.3720(4) | 4.3720(4) |
| **octahedral network ($OsO_6$ octahedra)** | | | | |
| O – O | [6] | 2.653(2) | - | - |
| O1 – O2 | [6] | - | 2.653(4) | 2.653(4) |
| **octahedral network ($O_6$ octahedra)** | | | | |
| O – O | [8] | 2.757(5) | - | - |
| O1 – O1 | [8] | - | 2.819(5) | 2.816(5) |
| O2 – O2 | [8] | - | 2.695(4) | 2.697(5) |

*[1] Numbers in square brackets indicate bond multiplicities.

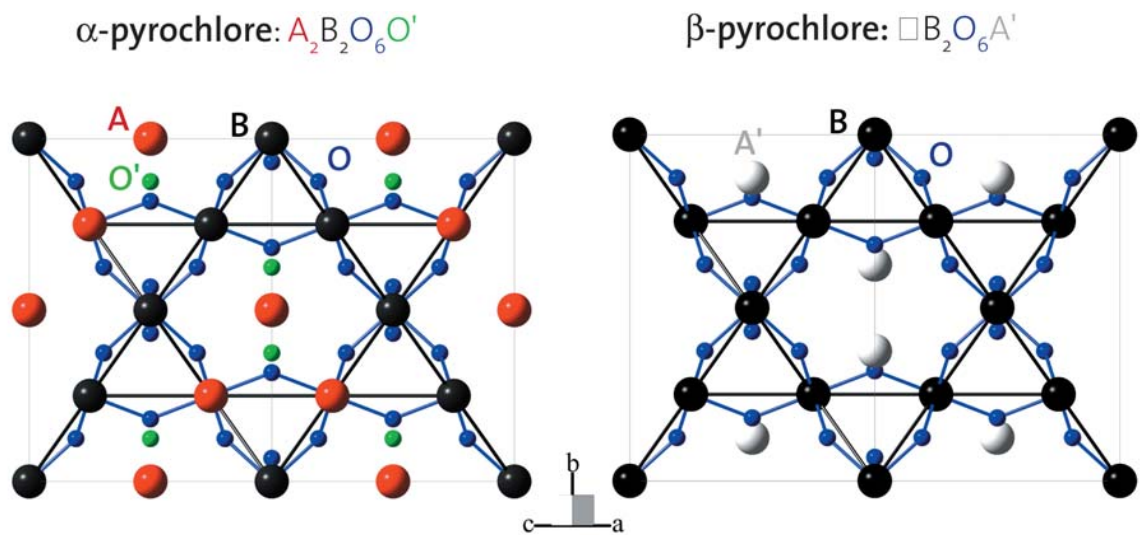

FIGURE 1 (color online). The α-pyrochlore (left) and β-pyrochlore (right) structures shown as a projection along [110]. The β-pyrochlore structure $\square_2B_2O_6A'$ (B = Os; A' = K, Rb, Cs) is derived form the α-pyrochlore structure $A_2B_2O_6O'$.

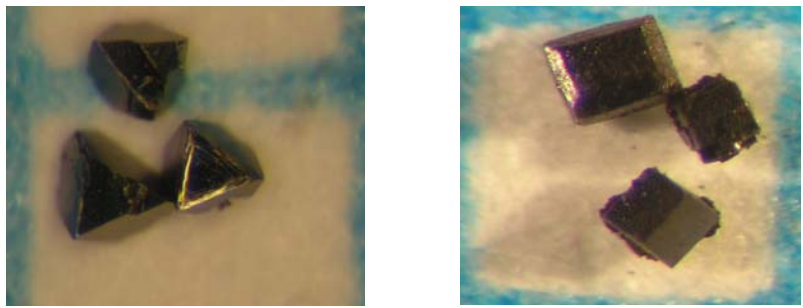

FIGURE 2 (color online). Single crystals of $KOs_2O_6$ grown in a quartz ampoule. The edge of the picture is about 1mm.

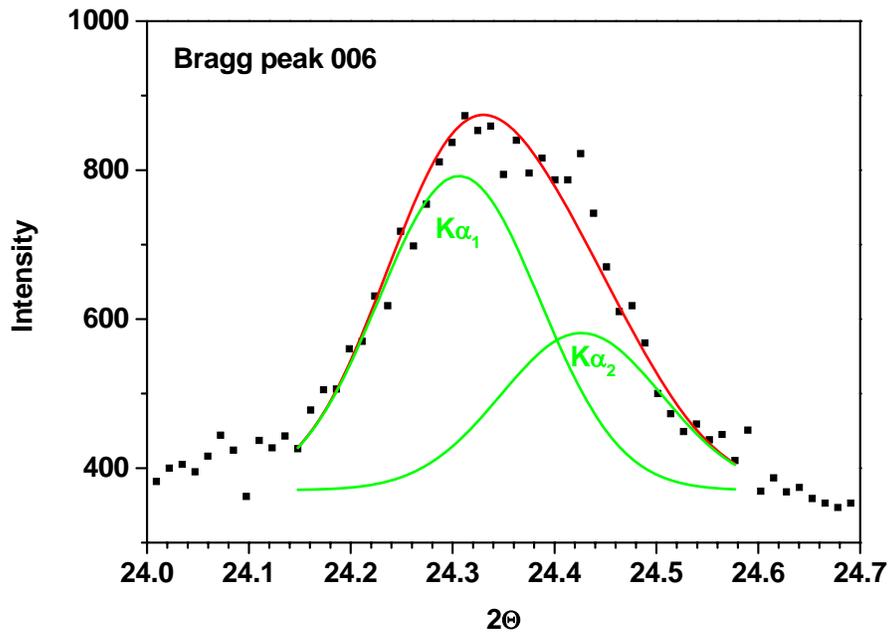

FIGURE 3 (color online). Reciprocal space scan of the (0 0 6) Bragg peak violating the condition of systematic absences for the $Fd\bar{3}m$ space group symmetry (00l: l ≠ 4n). The crystal of $KOs_2O_6$ was measured with the P4 diffractometer at room temperature (200 steps with 100 seconds per step, Gaussian fit of $K\alpha_1$ and $K\alpha_2$ with HWHM and amplitude constraint: $HWHM_{K\alpha1} = HWHM_{K\alpha2}$ and $A_{K\alpha2} = 0.5 \cdot A_{K\alpha1}$).

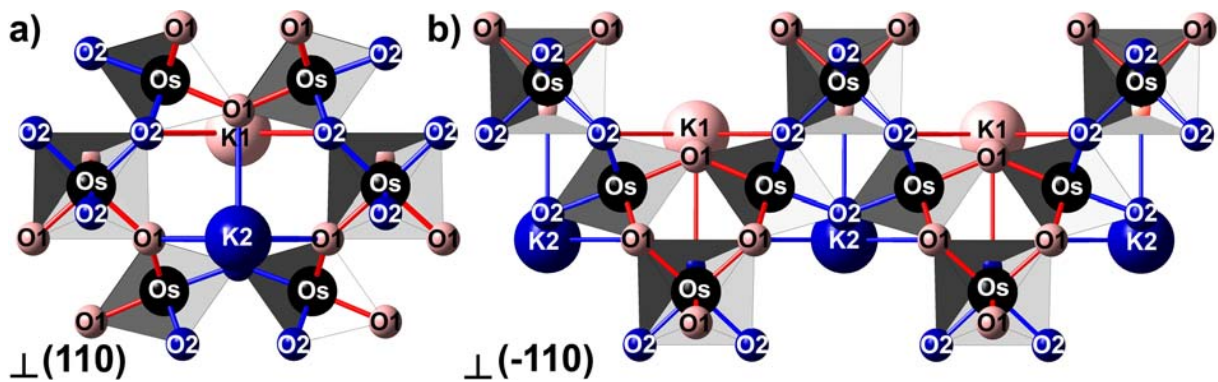

FIGURE 4 (color online). Structural details of β-pyrochlore $KOs_2O_6$ at room temperature with $F\bar{4}3m$ space group symmetry showing the anisotropic nature of the potassium channels. Interatomic distances O2-K1 = 3.141 Å and O1-K2 = 3.059 Å are longer and shorter, respectively, compared to the ideal β-pyrochlore structure with $Fd\bar{3}m$ symmetry (O-K = 3.099 Å). Shown is the structure projected along the [110] direction (a) and along the [-110] direction (b). Spheres represent K (large size), Os (medium size), and O (small size).

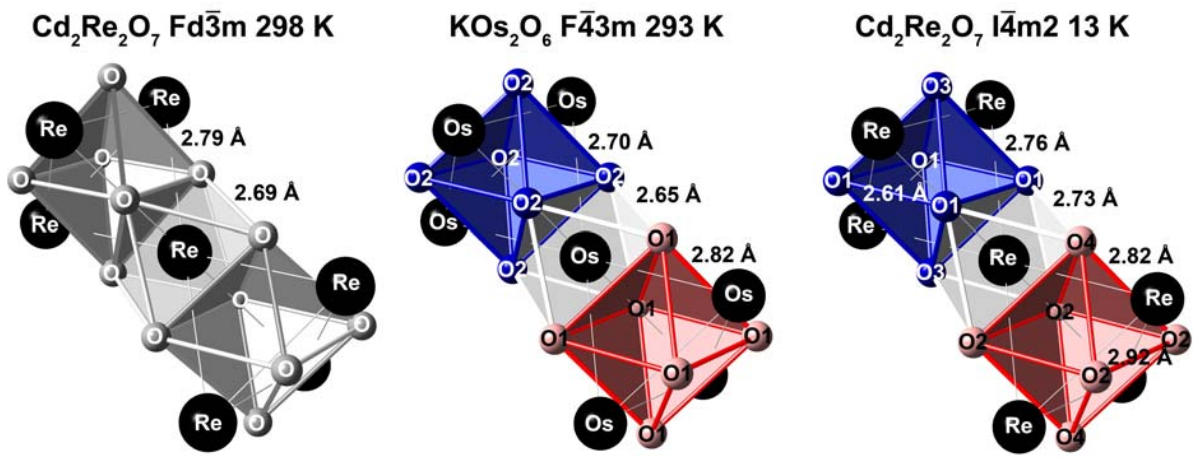

FIGURE 5 (color online). Structural details of β-pyrochlore $KOs_2O_6$ at room temperature with $F\bar{4}3m$ space group symmetry (middle) compared to $Cd_2Re_2O_7$ at 298K (left) and 13 K (right). Atomic parameters for $Cd_2Re_2O_7$ were taken from Weller et al..[17] For $KOs_2O_6$, the O-O interatomic distances are shown in the octahedral network: the length O1–O1 = 2.82 Å is longer and the length O2-O2 = 2.70 Å is shorter than the distance O-O = 2.76 Å for the ideal β-pyrochlore structure ($Fd\bar{3}m$). Large spheres represent Os, small spheres signify O.

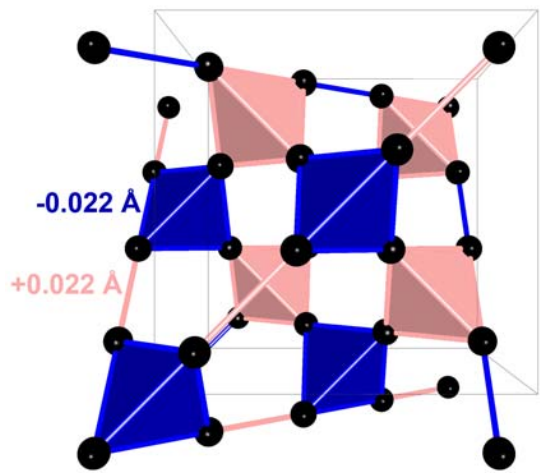

FIGURE 6 (color online). Tetrahedral network of the Os atoms (spheres) in β-pyrochlore $KOs_2O_6$ with $F\bar{4}3m$ space group symmetry. Two different Os tetrahedra are shown which exhibit breathing mode like behaviour: tetrahedra with increased Os-Os length (3.592 Å) and tetrahedra with decreased Os-Os length (3.548 Å), compared to the ideal β-pyrochlore structure with $Fd\bar{3}m$ symmetry (Os-Os = 3.570 Å).

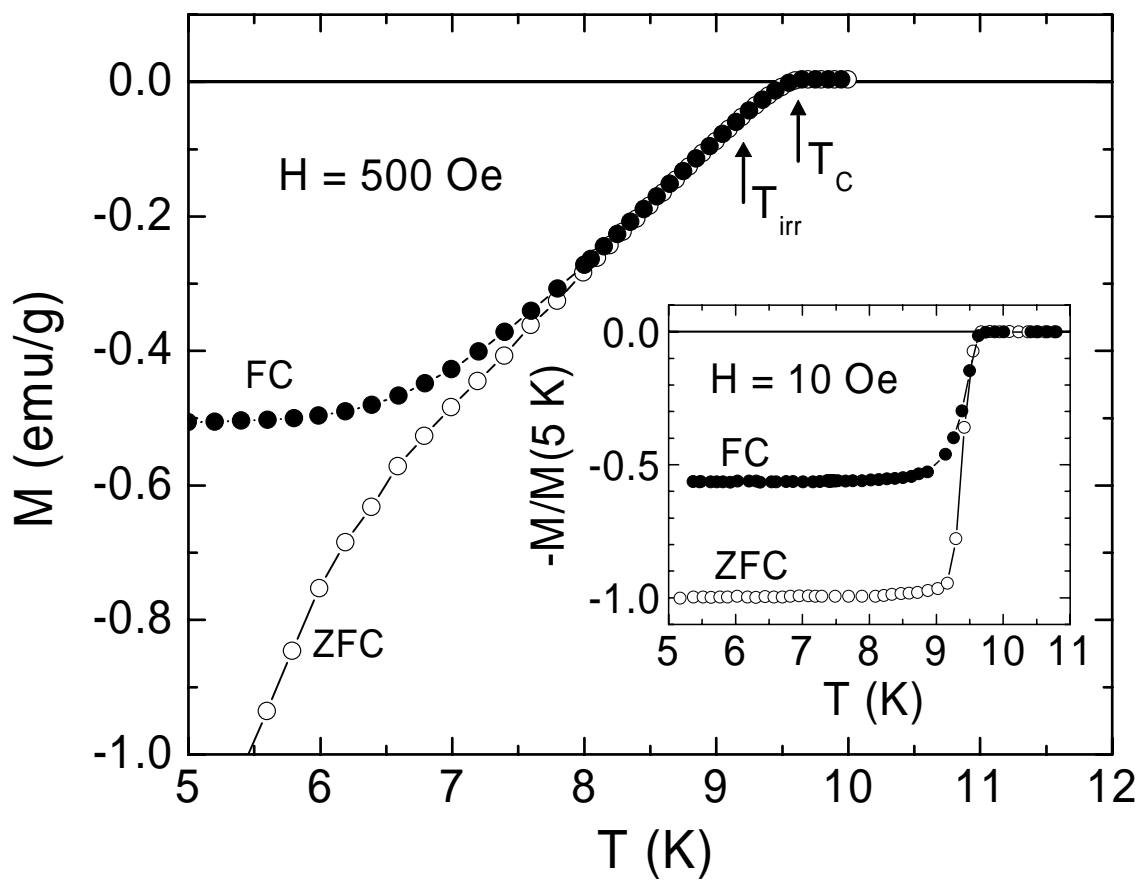

FIGURE 7. Magnetization (M) versus temperature at constant field H = 500 Oe for a set of 28 $KOs_2O_6$ crystals. The results are obtained upon heating from zero-field-cooled state (ZFC) and upon cooling (FC). The superconducting transition temperature $T_c$ and the irreversibility temperature $T_{irr}$ are marked by arrows. The inset shows M(T) reduced to M at 5 K for an individual $KOs_2O_6$ crystal in a field of 10 Oe.

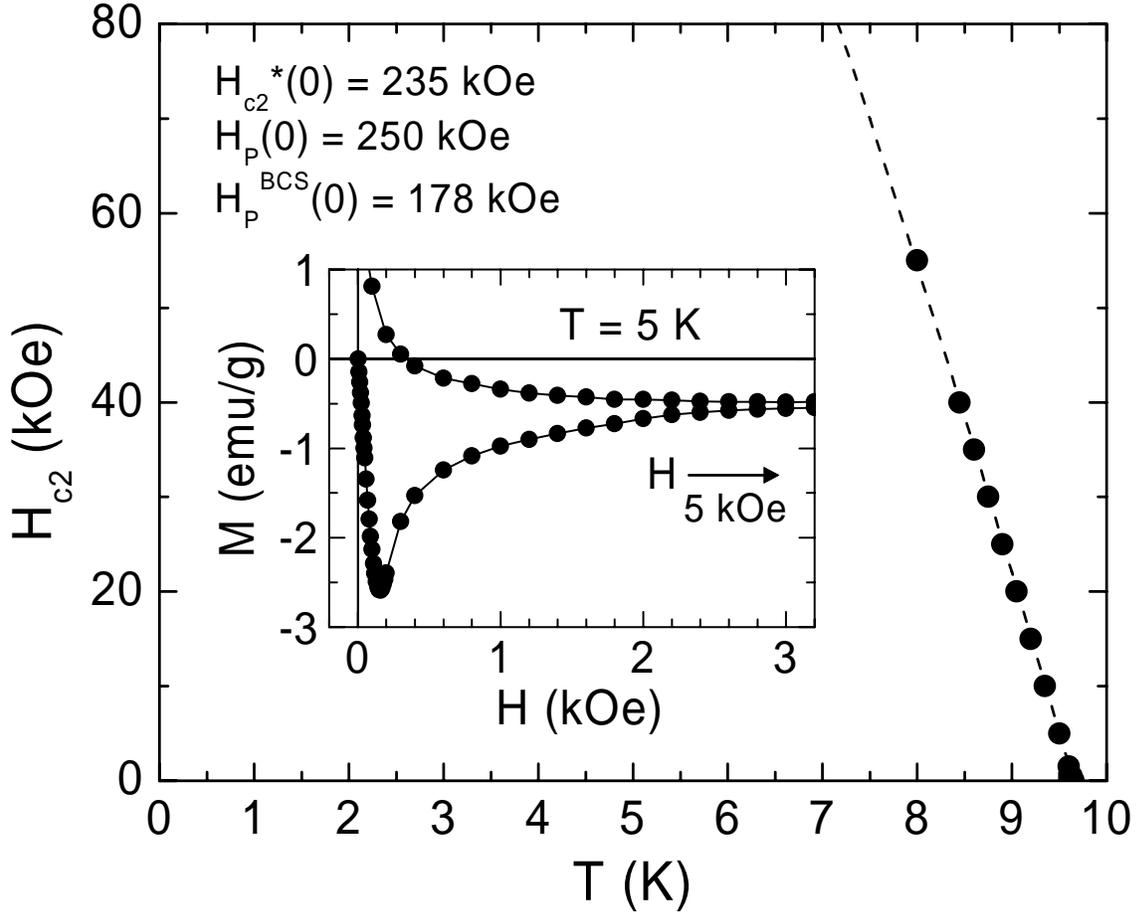

FIGURE 8. Upper critical field $H_{c2}$ close to the zero-field superconducting transition temperature $T_c$ = 9.65 K for the $KOs_2O_6$ crystals. The orbital critical field at zero temperature $H_{c2}^*(0)$ = 235 kOe, calculated within the GLAG theory (clean limit), is close to the Pauli limiting field $H_p(0) \cong 250$ kOe. The dashed line is a fit of experimental points to the formula $H_{c2}(T) = H_{c2}^*(0)(1-t^n)$ with n = 1.5, where t = $T/T_c$. The inset shows a low-field part of the virgin magnetization curve obtained at 5 K for H swept from zero to 5 kOe and then back to zero.